\documentstyle[epsfig]{prhep97}

\makeatletter
\let\chapter\hid@chapter
\makeatother

\def\beq{\begin{equation}}
\def\eeq{\end{equation}}
\def\beqar{\begin{eqnarray}}
\def\eeqar{\end{eqnarray}}
\def\barr#1{\begin{array}{#1}}
\def\earr{\end{array}}
\def\bfi{\begin{figure}}
\def\efi{\end{figure}}
\def\btab{\begin{table}}
\def\etab{\end{table}}
\def\bce{\begin{center}}
\def\ece{\end{center}}

\def\text{\textstyle}

\def\al{\alpha}

\def\De{\Delta}

\def\reffi#1{\mbox{Fig.~\ref{#1}}}

\def\citere#1{\mbox{Ref.~\cite{#1}}}

\newcommand{\GeV}{\unskip\,{\mathrm GeV}}
\newcommand{\MeV}{\unskip\,{\mathrm MeV}}

\def\mathswitchr#1{\relax\ifmmode{\mathrm{#1}}\else$\mathrm{#1}$\fi}

\newcommand{\PW}{\mathswitchr W}
\newcommand{\PZ}{\mathswitchr Z}

\newcommand{\PH}{\mathswitchr H}

\newcommand{\Pt}{\mathswitchr t}

\def\mathswitch#1{\relax\ifmmode#1\else$#1$\fi}

\newcommand{\MW}{\mathswitch {M_\PW}}

\newcommand{\MZ}{\mathswitch {M_\PZ}}
\newcommand{\MH}{\mathswitch {M_\PH}}

\newcommand{\Mt}{\mathswitch {m_\Pt}}

\newcommand{\GF}{\mathswitch {G_\mu}}

\newcommand{\fea}{{\em FeynArts}}

\newcommand{\two}{{\em TwoCalc}}

\begin{document}
%\pagenumbering{empty}

\thispagestyle{empty}

\null
\hfill KA-TP-24-1997\\
\null
\hfill hep-ph/9712227\\
\vskip .8cm
\begin{center}
{\Large \bf Sensitivity of Two-Loop Corrections to Muon Decay\\[.5em]
to the Higgs-Boson Mass%
\footnote{Contribution to the proceedings of the {\em International 
Europhysics Conference on High-Energy Physics}, Jerusalem, Israel, 
August 19--26, 1997.}
}
\vskip 2.5em
{\large
{\sc Georg Weiglein}\\[1ex]
{\normalsize \it Institut f\"ur Theoretische Physik, Universit\"at
Karlsruhe,
D-76128 Karlsruhe, Germany}
}
\vskip 2em
\end{center} \par
\vskip 1.2cm
\vfil
{\bf Abstract} \par
The Higgs-mass dependence of the two-loop contributions to muon decay
is analyzed at the two-loop level. Exact results are given for the
Higgs-dependent two-loop corrections associated with the fermions,
i.e.\ no expansion in the top-quark and the Higgs-boson mass is made.
The remaining theoretical uncertainties in the Higgs-mass
dependence of $\De r$ are discussed.
\par
\vskip 1cm
\null
\setcounter{page}{0}
\clearpage

% The following definitions need to be customised;

% Will appear on page headings
\authorrunning{G.\,Weiglein}
\titlerunning{{\talknumber}: Higgs-Mass Dependence of Two-Loop
Corrections to $\Delta r$}
 
% Now the full name of author and talk

% For plenary talks, the talk number is that of the session
\def\talknumber{710} 

\title{{\talknumber}: Sensitivity of Two-Loop Corrections to Muon Decay
to the Higgs-Boson Mass}
%Higgs-Mass Dependence of Two-Loop Corrections to $\Delta r$}
\author{Georg\,Weiglein
(georg@itpaxp5.physik.uni-karlsruhe.de)}
\institute{Institut f\"ur Theoretische Physik, Universit\"at Karlsruhe,
76128 Karlsruhe, Germany}

\maketitle

\begin{abstract}
The Higgs-mass dependence of the two-loop contributions to muon decay 
is analyzed at the two-loop level. Exact results are given for the
Higgs-dependent two-loop corrections associated with the fermions,
i.e.\ no expansion in the top-quark and the Higgs-boson mass is made.
The remaining theoretical uncertainties in the Higgs-mass
dependence of $\De r$ are discussed.
\end{abstract}
\section{Introduction}

The experimental accuracy meanwhile reached for the electroweak
precision observables allows to test the electroweak Standard
Model (SM) at its quantum level, where all parameters of the model
enter the theoretical predictions. In this way one is able to derive
constraints on the mass of the Higgs boson, which is the last missing
ingredient of the minimal SM. {}From the most recent global SM fits to
all available data one obtains an upper bound for the Higgs-boson mass
of $420$~GeV at $95\%$ C.L.~\cite{datasum97}. This bound is
considerably affected by the error in the theoretical
predictions due to missing higher-order corrections, which gives rise
to an uncertainty of the upper bound of about $100$~GeV.
The main uncertainty in this context comes from the electroweak
two-loop corrections, for which the results obtained so far have
been restricted to expansions for asymptotically large values of the
top-quark mass, $\Mt$, or the Higgs-boson mass,
$\MH$~\cite{twoloopres,gamb}.

In order to improve this situation, an exact evaluation of
electroweak two-loop contributions would be desirable, where no
expansion in $\Mt$ or $\MH$ is made. In this paper the Higgs-mass
dependence of the two-loop contributions to $\De r$ in the SM is 
studied~\cite{sbaugw2}. Exact results for the corrections associated 
with the fermions are presented.

%\section{Method of calculation}
\section{Higgs-mass dependence of $\De r$}

The relation between the vector-boson masses in terms of the Fermi
constant $\GF$ reads~\cite{dr1loop}
\beq
\MW^2 \left(1 - \frac{\MW^2}{\MZ^2}\right) =
\frac{\pi \al}{\sqrt{2} \GF} \left(1 + \De r\right),
\eeq
where the radiative corrections are contained in the quantity $\De r$.
In the context of this paper we treat $\De r$ without resummations,
i.e.\ as being fully expanded up to two-loop order,
%\beq
$
\De r = \De r_{(1)} + \De r_{(2)} + %{\cal O}(\al^3) .
\ldots \; .
$
%\eeq
The theoretical predictions for $\De r$ are obtained by calculating
radiative corrections to muon decay.
We study the variation of the two-loop contributions to $\De r$
with the Higgs-boson mass by considering the subtracted quantity
\beq
\label{eq:DeltaRsubtr}
\De r_{(2), {\mathrm subtr}}(\MH) =
\De r_{(2)}(\MH) - \De r_{(2)}(\MH = 65\GeV),
\eeq
where $\De r_{(2)}(\MH)$ denotes the two-loop contribution to
$\De r$. Potentially large $\MH$-dependent contributions are the
corrections associated with the top quark, since the Yukawa coupling of
the Higgs boson to the top quark is proportional to $\Mt$, and the
contributions which are proportional to $\De\al$. 

The methods used for the calculations discussed in this
paper have been outlined in \citere{sbaugw1}. The generation of the
diagrams and counterterm contributions is done with the help of the
computer-algebra program \fea\ \cite{fea}. Making use of two-loop
tensor-integral decompositions, the generated amplitudes are reduced to
a minimal set of standard scalar integrals with the program 
\two~\cite{two}. The renormalization is performed within the complete 
on-shell scheme~\cite{onshell}, i.e.\ physical parameters are used 
throughout. The two-loop scalar integrals are evaluated numerically 
with one-dimensional integral representations~\cite{intnum}. These 
allow a very fast calculation of the integrals with high precision 
without any approximation in the masses.

%\section{Results for the Higgs-mass dependence}

%We begin with the Higgs-mass dependence of the two-loop top-quark
%contributions and consider the quantity $\De r^{\mathrm top}_{(2),
%{\mathrm subtr}}(\MH)$, which denotes the contribution of the top/bottom
%doublet to $\De r_{(2), {\mathrm subtr}}(\MH)$. From the one-particle
%irreducible diagrams obviously those graphs contribute to $\De
%r^{\mathrm top}_{(2), {\mathrm subtr}}$ that contain both the top
%and/or bottom quark and the Higgs boson. It is easy to see that
%only two-point functions enter in this case, since all graphs where 
%the Higgs boson couples to the
%muon or the electron may safely be neglected. Although no two-loop
%three-point function enters, there is nevertheless a contribution from
%the two-loop and one-loop vertex counterterms. If the field
%renormalization constants of the W~boson are included (which cancel in
%the complete result), the vertex counterterms are separately finite.
We first consider the contribution of the top/bottom doublet, which is
denoted as $\De r^{\mathrm top}_{(2), {\mathrm subtr}}(\MH)$.
From the one-particle irreducible diagrams obviously those graphs
contribute to $\De r^{\mathrm top}_{(2), {\mathrm subtr}}$ that contain
both the top and/or bottom quark and the Higgs boson. 
The technically most complicated contributions arise from the mass and
mixing-angle renormalization. Since it is performed in the on-shell
scheme, the evaluation of the W- and Z-boson self-energies is required
at non-zero momentum transfer.

The contribution of the terms proportional to $\De \al$ has the simple
form $\De r^{\De\al}_{(2), {\mathrm subtr}}(\MH) = 2 \De\al \, \De
r_{(1), {\mathrm subtr}}(\MH)$ and can easily be obtained by a proper
resummation of one-loop terms~\cite{sirresum}. 
The remaining fermionic contribution,
$\De r^{\mathrm lf}_{(2), {\mathrm subtr}}$,
is the one of the light fermions,
i.e.\ of the leptons and of the quark doublets of the first and second
generation,
which is not contained in $\De\al$. Its structure is analogous to
$\De r^{\mathrm top}_{(2), {\mathrm subtr}}$,
but because of the negligible coupling of
the light fermions to the Higgs boson much less diagrams contribute.

\begin{figure}[ht]
\begin{center}
%\mbox{}\\[0.2cm]
\mbox{
\psfig{figure=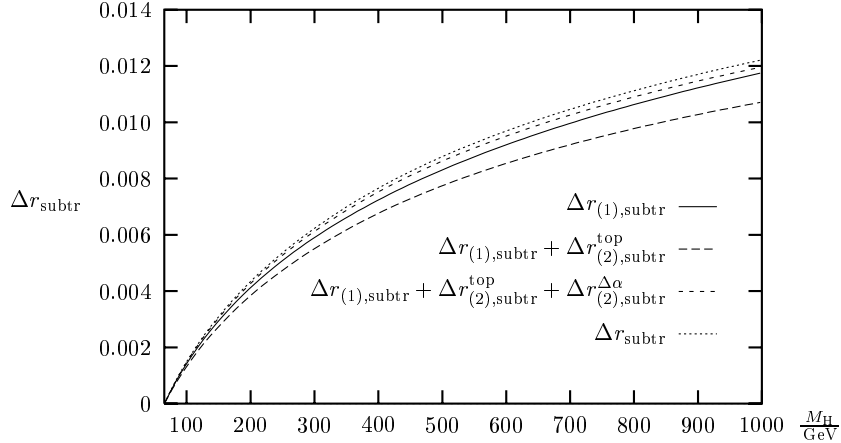,%width=9.5cm,height=6.5cm,
              width=9.5cm,
              bbllx=142pt,bblly=445pt,bburx=480pt,bbury=633pt}}
%\\[0.8cm]
\caption{One-loop and two-loop contributions to $\Delta r$
subtracted at $\MH=65\,\GeV$.
$\De r_{\mathrm subtr}$ is the result for the full one-loop and
fermionic
two-loop contributions to $\De r$, as defined in the text.
\label{fig:delr2}
}
\end{center}
\end{figure}

The total result for the one-loop and fermionic two-loop contributions
to $\De r$, subtracted at $\MH=65\,\GeV$, reads
\beq
\De r_{\mathrm subtr} \equiv \De r_{(1), {\mathrm subtr}} +
\De r^{\mathrm top}_{(2), {\mathrm subtr}} +
\De r^{\De\al}_{(2), {\mathrm subtr}} +
\De r^{\mathrm lf}_{(2), {\mathrm subtr}} .
\eeq
It is shown in \reffi{fig:delr2}, where separately also
the one-loop contribution
$\De r_{(1), {\mathrm subtr}}$, as well as
$\De r_{(1), {\mathrm subtr}} + \De r^{\mathrm top}_{(2), {\mathrm
subtr}}$, and
$\De r_{(1), {\mathrm subtr}} + \De r^{\mathrm top}_{(2), {\mathrm
subtr}} + \De r^{\De\al}_{(2), {\mathrm subtr}}$
are shown for $\Mt = 175.6 \GeV$.
The two-loop contributions $\De r^{\mathrm top}_{(2), {\mathrm
subtr}}(\MH)$ and $\De r^{\De\al}_{(2), {\mathrm subtr}}(\MH)$ turn out
to be of similar size and to cancel each other to a large extent.
In total, the inclusion of the higher-order contributions discussed
here
leads to a slight increase in the sensitivity to the Higgs-boson mass
compared to the pure one-loop result.

We have compared the result for $\De r^{\mathrm top}_{(2), {\mathrm
subtr}}$ with the result obtained via an expansion in $\Mt$ up to
next-to-leading order, i.e.\ ${\cal O}(\GF^2 \Mt^2 \MZ^2)$~\cite{gamb}.
The results agree within about $30 \%$ of $\De r^{\mathrm top}_{(2),
{\mathrm subtr}}(\MH)$, which amounts to a difference in $\MW$ of up 
to about $4$~MeV~\cite{sbaugw2}.

%In \refta{tab:DeltaMW} the shift in $\MW$ corresponding to $\De
%r_{\mathrm
%subtr}(\MH)$, i.e.~the change in the theoretical prediction for $\MW$
%when varying the Higgs-boson mass from $65\, \GeV$ to $1\, \TeV$, is
%shown for three values of the top-quark mass, $\Mt = 170, 175, 180\,
%\GeV$. The dependence on the precise value of $\Mt$ is rather mild,
%which is expected from the fact that $\Mt$ enters here only at the
%two-loop level and that $\De r^{\mathrm top}_{(2), {\mathrm
%subtr}}(\MH)$ has a local maximum in the region of $\Mt = 175$~GeV
%(see \citere{sbaugw2}).
%
%\btab
%$$
%\barr{|c||c|c|c|c|c|c|c|c|} \hline
%\MH/\GeV & 65 & 100 & 200 & 300 & 400 & 500 & 600 & 1000 \\ \hline
%\hline
%\Delta \MW(\MH)/\MeV,\; \Mt=170\,\GeV &
%0 & -22.6 & -65.8 & -94.5 & -116 & -133 & -147 & -185  \\ \hline
%\Delta \MW(\MH)/\MeV,\; \Mt=175\,\GeV &
%0 & -22.8 & -66.3 & -95.2 & -117 & -134 & -148 & -187  \\ \hline
%\Delta \MW(\MH)/ \MeV,\; \Mt=180\,\GeV &
%0 & -23.0 & -66.8 & -96.0 & -118 & -135 & -149 & -188  \\ \hline
%\earr
%$$
%\caption{The shift in the %dependence of the %shift in $\MW$ on the
%theoretical prediction for $\MW$ caused by varying the Higgs-boson mass
%in the interval $65\, \GeV \leq \MH \leq 1\, \TeV$
%for three values of $\Mt$.
%\label{tab:DeltaMW}}
%\etab
%

Regarding the remaining Higgs-mass dependence of $\De r$ at the
two-loop level, there are only purely bosonic corrections left, which
contain no specific source of enhancement. They
can be expected to yield a contribution to
$\De r_{(2), {\mathrm subtr}}(\MH)$ of about the same size as
$\left.\left(\De r^{\mathrm bos}_{(1)}(\MH)\right)^2\right|_{\mathrm
subtr}$, where $\De r^{\mathrm bos}_{(1)}$ denotes the bosonic
contribution to $\De r$ at the one-loop level. The contribution of
$\left.\left(\De r^{\mathrm bos}_{(1)}(\MH)\right)^2\right|_{\mathrm
subtr}$ amounts to only about $10 \%$ of $\De r^{\mathrm top}_{(2),
{\mathrm subtr}}(\MH)$ corresponding to a shift of about $2 \MeV$ in the
W-boson mass.
This estimate agrees well with the values obtained for the
Higgs-mass dependence from the formula in \citere{hakni} for the leading 
term proportional to $\MH^2$ in an
asymptotic expansion for large Higgs-boson mass.
The Higgs-mass dependence of the term proportional to $\MH^2$ amounts to
less than $15 \%$ of $\De r^{\mathrm top}_{(2), {\mathrm subtr}}(\MH)$
for reasonable values of $\MH$.

\section{Conclusions}

We have analyzed the Higgs-mass dependence of the
relation between the gauge-boson masses at the two-loop level
by considering the subtracted quantity $\De r_{\mathrm subtr}(\MH) = \De
r(\MH) - \De r(\MH = 65\, \GeV)$. Exact results have been presented for
the fermionic contributions, i.e.\ no expansion in the top-quark and the
Higgs-boson mass has been made. The extra shift coming from the
purely bosonic two-loop corrections has been estimated to be relatively
small. Considering the envisaged
experimental error of $\MW$ from the measurements at LEP2 and the
Tevatron of $\sim 20 \MeV$, we conclude that
the theoretical uncertainties due to
unknown higher-order corrections in the Higgs-mass dependence of
$\De r$ are now under control.

\smallskip

The author thanks S.~Bauberger for fruitful collaboration on this
subject.

% ---- Bibliography ----
%

\end{document}